# THE NUMI NEUTRINO BEAM AT FERMILAB*

S. Kopp,[#] Department of Physics, University of Texas, Austin, TX 78712, U.S.A.

*Abstract*

The Neutrinos at the Main Injector (NuMI) facility at Fermilab began operations in late 2004. NuMI will deliver an intense $\nu_\mu$ beam of variable energy (2-20 GeV) directed into the Earth at 58 mrad for short (~1km) and long (~700-900 km) baseline experiments. Several aspects of the design and results from early commissioning runs are reviewed.

## INTRODUCTION

A new beam line facility, the "Neutrinos at the Main Injector" (NuMI) [1], has been constructed at the Fermi National Accelerator Laboratory in Illinois, to deliver an intense $\nu_\mu$ beam to what is planned to be a variety of experiments. The first experiment, MINOS [2], will perform definitive spectrum measurements which demonstrate the effect of $\nu$ oscillations. The second, MINERνA [3], is an experiment 1 km from the NuMI target to perform neutrino cross section measurements. A third proposal, NOνA [4], has been approved to explore the phenomenon of CP violation in neutrinos.

NuMI is a tertiary beam resulting from the decays of pion and kaon secondaries produced in the NuMI target. Protons of 120 GeV are fast-extracted (spill duration 8.6 μsec) from the Main Injector (MI) accelerator and bent downward by 58 mrad toward Soudan, MN (see Figure 1). The beam line is designed to accept $4\times10^{13}$ protons per pulse (ppp). The repetition rate is 0.53 Hz, giving ~$4\times10^{20}$ protons on target per year.

## MAIN INJECTOR BEAM

The MI is fed multiple batches from the 8 GeV Booster accelerator, of which 5 are extracted to NuMI. The Booster can deliver $5\times10^{12}$ *p*/batch, and efforts are underway to increase this number [5]. Studies of running multiple batches simultaneously in the MI have achieved accelerated beams of up to $3\times10^{13}$ protons. This number is improving because of the installation and commissioning of a new digital damper system [6] and a beam loading compensation system [7] for the MI.

Initiatives to produce higher intensity beam from the MI with the existing proton source, including multiple batch "stacking" of more than 6 Booster batches, could increase the beam intensity to NuMI by a factor of 1.5. Already, synchronized transfer of 2 batches from Booster to the MI (this is referred to as batch "cogging"[8]), followed by slip-stacking of the batches to a single batch of ~$7\times10^{12}$ protons[9], has been in use for production of antiprotons since August, 2004. The goal in 2005 is to continue slip-stacking for p-bar production, and accelerate $3.3\times10^{13}$ protons per MI cycle ($8\times10^{12}$ ppp slip-stacked for antiproton production, $2.5\times10^{13}$ ppp for NuMI). Future studies will address the feasibility of slip-stacking of additional batches for NuMI.

## PRIMARY BEAM TRANSPORT

The MI beam is extracted by a set of three kickers and three Lambertsons at the MI60 region of the MI. The transport line to the NuMI target is 350m long. The transport line will maintain losses below $10^{-5}$ to reduce component activation and to reduce activation of ground water. In anticipation of batch stacking in the MI, the momentum acceptance of the transport line is $\Delta p/p = 0.0038$ at $40\pi$ mm-mr. Injection errors of ~1 mm lead to targeting errors of ~0.5 mm. The beam line has 2 toroids, 44 loss monitors, 24 BPM's, 19 dipole correctors, and 10 retractable segmented foil secondary emission detectors for measuring beam profile and halo [10].

## TARGET AND HORNS

The primary beam is focused onto a graphite production target of $6.4\times15\times940$ mm$^3$, segmented longitudinally into 47 fins. The beam size at the target is 1 mm. The target is water cooled via stainless steel lines at the top and bottom of each fin and is contained in an aluminum vacuum can with beryllium windows. It is electrically isolated so it can be read out as a Budal monitor [12]. The target has a safety factor of about 2.2 for the fatigue lifetime of $10^7$ pulses (1 NuMI year) given the calculated dynamic stress of $4\times10^{13}$ protons/pulse and 1 mm spot size. A prototype target was tested in the Main Injector in 1998 at peak energy densities exceeding that expected in NuMI [11]. Studies indicate that the existing NuMI target could withstand up to a 1 MW proton beam if the beam spot size is increased from 1 mm to 2-3 mm [13].

The particles produced in the target are focused by two magnetic "horns" [14]. The 200 kA peak current produces a maximum 30 kG toroidal field which sign- and momentum-selects the particles from the target. Field measurements on a horn prototype show the expected $1/r$ fall-off to within a percent. The horns are designed to withstand $10^7$ pulses (1 NuMI year), and tests without beam of the prototype horn have so far achieved this. The relative placement of the two horns and the target optimizes the momentum focus for pions, hence the peak neutrino beam energy. To fine-tune the beam energy, the target is mounted on a rail-drive system with 2.5 m of longitudinal travel, permitting remote change of the beam energy without accessing the horns and target [15]. The neutrino spectra from several target position settings are shown in Figure 2. Both the target and horns are protected by an upstream collimating baffle, consisting of air-cooled graphite with an inner 11mmØ bore.

*Work supported by U.S. DoE, contracts DE-FG03-93ER40757 and DE-AC02-76CH3000
[#]kopp@hep.utexas.edu

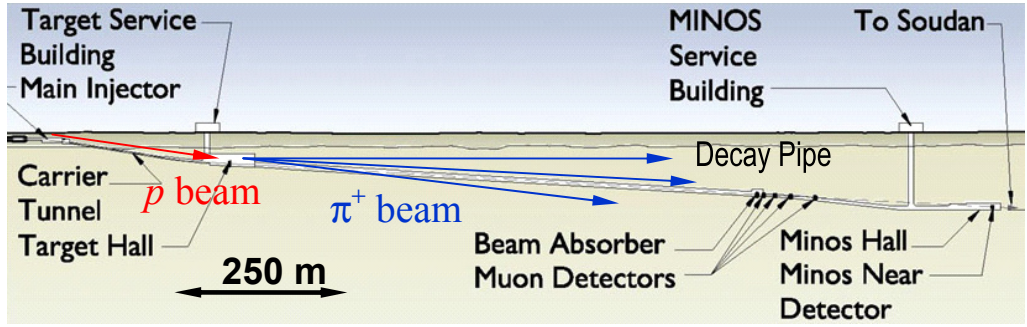

Figure 1: Elevation view of the NuMI beam, indicating primary beam transport, target station, decay volume, beam absorber, muon detectors, and detector hall housing the first MINOS detector.

Each of the two horns and the target are supported beneath shielding modules which are lowered by overhead crane into the target cavern. Using the modules, failed horns or targets may be lowered remotely into a shielding pit for disposal. The horns, target, and support modules are cooled in the target cavern via recirculating air system flowing at 25 km/hr. Should significant proton intensity upgrades be pursued, this air system would require upgrade and the air flow volume better sealed to contain contamination from radioactive air molecules.

## DECAY VOLUME AND ABSORBER

Particles are focused by the horns into a 675 m long, 2 m diameter steel pipe evacuated to ~1 Torr. This length is approximately the decay length of a 10 GeV pion. The entrance window to the decay volume is a ellipsoidal bell-shaped steel window 1.8 cm thick, with a 1 mm thick aluminum window 1 m in diameter at its center where 95% of the entering pions traverse. The decay volume is surrounded by 2.5-3.5 m of concrete. Water cooling lines around the exterior of the decay pipe remove the 150 kW of beam heating deposited in the steel pipe and concrete.

Earlier plans to instrument the decay volume with a current-carrying wire have been abandoned due to budget constraints. Such a device provides a toroidal field that continuously focuses pions along the decay pipe length and increases the ν flux by approximately 30% [16].

At the end of the decay volume is a beam absorber consisting of a $1.2 \times 1.2 \times 2.4$ m$^3$ water-cooled aluminum core, a 1 m layer of steel blocks surrounding the core, followed by a 1.5 m layer of concrete blocks. The core absorbs 65 kW of beam power, but can sustain the full 400kW beam power for up to an hour in the event of mistargeting. In the event of a proton intensity upgrade the core would require no modification, but the steel blocks might require cooling.

## SECONDARY AND TERTIARY BEAM INSTRUMENTATION

Ionization chambers are used to monitor the secondary and tertiary particle beams [17]. An array is located immediately upstream of the absorber, as well as at three muon "pits", one downstream of the absorber, one after 8 m of rock, and a third after an additional 12 m of rock. These chambers monitor the remnant hadrons at the end of the decay pipe, as well as the tertiary muons from π and K decays. When the beam is tuned to the medium energy configuration, the pointing accuracy of the muon stations can align the neutrino beam direction to approximately 50 μradians in one spill. In NuMI, the hadron (muon) monitor will be exposed to charged particle fluxes of $10^9$/cm$^2$/spill, ($10^7$/cm$^2$/spill). Beam tests of these chambers indicate an order of magnitude safety factor in particle flux over the rates expected in NuMI before space charge buildup affects their operation.

## COMMISSIONING EXPERIENCE

Three short runs of the NuMI facility have transpired, each with a goal of commissioning additional components in the beam line. The first, in December, 2004, extracted $3 \times 10^{11}$ ppp from the MI, delivered it down the primary transport line and sent it to the NuMI beam absorber. No target was in place at this time. On the first pulse, beam was successfully extracted and on the 10$^{th}$ pulse (shown in Figure 3), beam was observed at the center of the Hadron Monitor detector, located just upstream of the absorber, thus verifying proper transport and alignment.

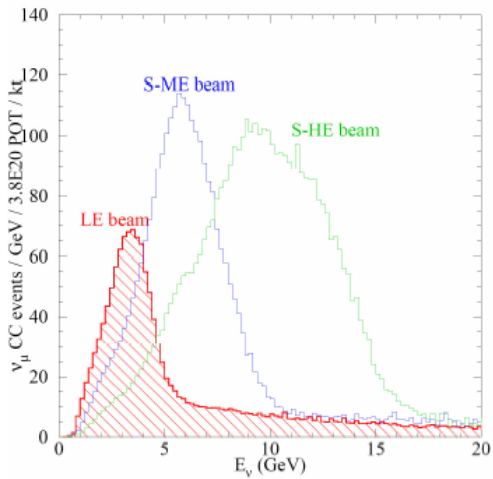

FIGURE 2: Neutrino beam energy spectra achieved at the MINOS experiment with the target in its nominal position inside the horn (LE), or retracted 1 m (ME) or 2.5 m (HE).

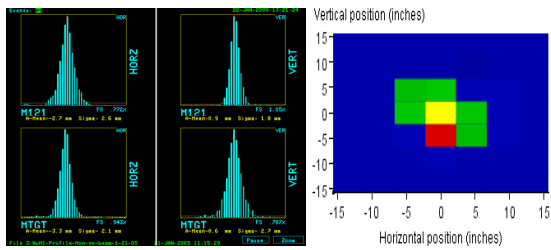

Figure 3: 1-dimensional proton beam profiles (horizontal and vertical) as seen in the last two foil SEM's upstream of the target hall as well as in the 2-dimensional Hadron Monitor array (between the decay volume and beam absorber).

The 2nd commissioning run in Jan., 2005, delivered $3\times10^{12}$ ppp onto the NuMI target and the energized horns focused the resulting secondaries into the decay volume. Neutrino interactions were observed in the first MINOS detector. As shown in Figure 4, the neutrino beam instrumentation was able to track the expected flux and energy of the neutrino beam being produced: a muon beam was observed whose intensity increased as the horns were ramped up and whose penetrating power (energy) increased as the target was moved to a higher neutrino-energy position, in accordance with expectations. As shown in Figure 5, beam-based alignment of the target and horns can be accomplished by using the secondary beam instrumentation to monitor particle fluences as the beam was scanned across the target, baffle, and horns.

The third run in March, 2005, commissioned the simultaneous acceleration of 120 GeV protons for NuMI and for producing antiprotons. The MI accelerated 7 Booster batches, 2 slip-stacked together for antiprotons, and 5 extracted to NuMI. A cycle time of 3 seconds and a per-pulse intensity of $2.3\times10^{13}$ protons were separately demonstrated. Neutrino interactions were observed in the MINOS detector in Soudan, MN, after approximately $7\times10^{17}$ protons on target, consistent with expectations.

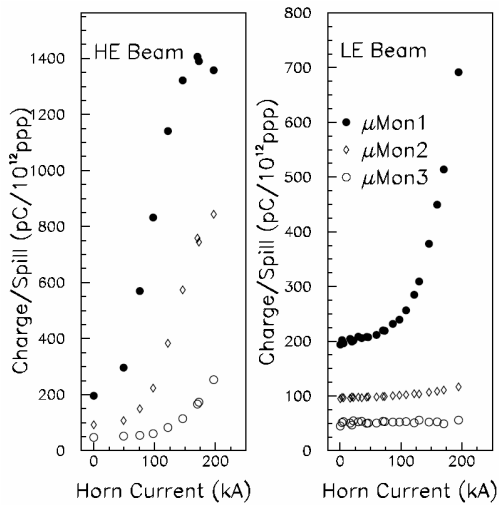

Figure 4: Total charge observed in the 3 muon monitor detectors as a function of the current in the focusing horns during a high energy (left) and low energy (right) neutrino run.

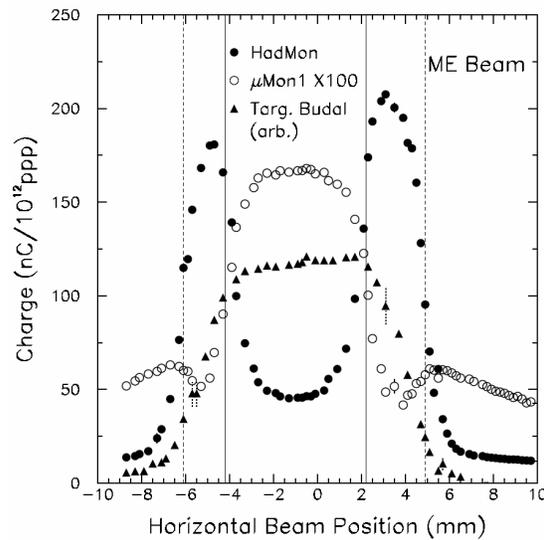

Figure 5: Signals from the Hadron Monitor, 1st Muon Monitor, and the target Budal monitor as the beam is scanned horizontally across the NuMI target. The solid (dashed) line represents the inferred outer (inner) edges of the target (collimating baffle).

## OUTLOOK

After a 4 week pause to repair a leak in the target cooling system, the NuMI facility resumed operations in April. In the coming year it is hoped that the Booster/MI complex will routinely deliver $2.5\times10^{13}$ ppp every 2sec. for NuMI, yielding enough data for preliminary results from the MINOS experiment by year's end.

## ACKNOWLEDGEMENTS

It is a pleasure to acknowledge the many who have contributed to the NuMI facility from the Fermilab Accelerator, Technical, and Particle Physics Divisions, as well as from the MINOS Collaboration.